\documentclass[
 reprint,
 nofootinbib,
 amsmath,amssymb,
 aps,
 prb,
]{revtex4-1}

\usepackage{graphicx}
\usepackage{xcolor}
\usepackage{transparent}
\usepackage{dcolumn}
\usepackage{bm}	
\usepackage{amssymb}
\usepackage{braket}
\usepackage{hyperref}
\usepackage{mathtools}
\usepackage{setspace}
\usepackage{amsmath}
\usepackage{esdiff}
\usepackage[sort&compress]{natbib}
\usepackage{caption}
\usepackage{subcaption}
\usepackage{amsthm}
\usepackage{fancyhdr}
\usepackage{titlesec}

\usepackage{graphicx}
\usepackage{xcolor}
\usepackage{bm}	
\usepackage{upgreek}
\usepackage{braket}
\usepackage{hyperref}
\usepackage{algorithm}
\usepackage{algpseudocode}
\usepackage{soul}

\DeclareRobustCommand*{\citen}[1]{%
  \begingroup
    \romannumeral-`\x 
    \setcitestyle{numbers}%
    \cite{#1}%
  \endgroup   
}

\begin{document}
\title{Extending qubit coherence by adaptive quantum environment learning}
\author{Eleanor Scerri}
\email{ds135@hw.ac.uk}
\address{SUPA, Institute of Photonics and Quantum Sciences, Heriot-Watt University, David Brewster Building, Edinburgh, EH14 4AS, UK}
\author{Erik M. Gauger}
\address{SUPA, Institute of Photonics and Quantum Sciences, Heriot-Watt University, David Brewster Building, Edinburgh, EH14 4AS, UK}
\author{Cristian Bonato}
\email{c.bonato@hw.ac.uk}
\address{SUPA, Institute of Photonics and Quantum Sciences, Heriot-Watt University, David Brewster Building, Edinburgh, EH14 4AS, UK}

\date{\today}

\begin{abstract}
Decoherence, resulting from unwanted interaction between a qubit and its environment, poses a serious challenge towards the development of quantum technologies. Recently, researchers have started analysing how real-time Hamiltonian learning approaches, based on estimating the qubit state faster than the environmental fluctuations, can be used to counteract decoherence. In this work, we investigate how the back-action of the quantum measurements used in the learning process can be harnessed to extend qubit coherence. We propose an adaptive protocol that, by learning the qubit environment, narrows down the distribution of possible environment states. While the outcomes of quantum measurements are random, we show that real-time adaptation of measurement settings (based on previous outcomes) allows a deterministic decrease of the width of the bath distribution, and hence an increase of the qubit coherence. We numerically simulate the performance of the protocol for the electronic spin of a nitrogen-vacancy centre in diamond subject to a dilute bath of $^{13}$C nuclear spin, finding a considerable improvement over the performance of non-adaptive strategies.
\end{abstract}

\maketitle

\section{Introduction}

Individual spins are an established platform for developing solid-state quantum technologies for improved metrology \cite{Balasubramanian2008, Togan2011, Grinolds2013, Soykal2017, Dolde2011, Lo2015, Thie2019, Schlussel2018, Gross2017, Abobeih2019, Zopes2018}, communication \cite{Kalb2017, Humphreys2018} and information processing \cite{Weber8513, Veldhorst2014, Veldhorst2017, Watson2018}. All quantum applications rely on the capability to preserve the coherence of quantum states. Due to the solid-state environment, coherence is only preserved on a finite timescale dictated by fluctuations originating from neighbouring impurities and spins. While coherence could be extended by minimising the concentration of impurities \cite{Balasubramanian2009, Nagy2019},  improving material quality is typically not straightforward and requires a considerable investment. Another possibility is to use quantum error correction methods \cite{Unden2016, Taminiau2014, Waldherr2014, Cramer2016} to preserve the quantum state of the system. However, this adds a significant overhead in terms of additional qubits required. 
For these reasons, the most widely adopted current mitigation strategy builds on pulse sequences such as dynamical decoupling \cite{Viola1999, Lange2010, Suter2016}, which can efficiently isolate the qubit from its environment at specific times during its evolution. This strategy, however, poses problems when the qubit is required on-demand and not at pre-determined intervals.
%

A different approach to extending qubit coherence relies on tracking and compensating environmental fluctuations through Hamiltonian learning. Recent significant advances in computational power and high-speed programmable electronics have made real-time learning algorithms experimentally viable \cite{Granade2012, Wang2017, Bonato2017, Santagati2019}. For example, by measuring the qubit faster than the fluctuations responsible for decoherence, one can adaptively compensate any detuning and maintain coherence \cite{Stepanenko2006}. This idea has been experimentally implemented  to increase the coherence time $T^*_2$ time for a spin in a quantum dot \cite{Shulman2014}, from the nanosecond to the microsecond timescale. For a single electronic spin in a GaAs quantum dot, interaction with the nuclear spin environment can be treated as a semi-classical time-varying magnetic field, due to the size of the spin bath.

In materials such as diamond, silicon or silicon carbide, the concentration of nuclear spins is much smaller than in GaAs, and a central spin only interacts with a dilute environment of a few nuclear spins. In this case, the semi-classical approximation is no longer viable and the full quantum dynamics of individual nuclear spins comes into play. Recent work has shown that by applying controlled pulse sequences to an electronic spin in diamond (1.1$\%$ $^{13}$C concentration), a dozen individual nuclear spins  can be detected \cite{Neumann2010, Robledo2011, Zhao2012, Zopes2018, Taminiau2019} and coherently controlled \cite{Taminiau2012, Unden2018}, opening the way to exploit the quantum properties of the bath. 
In this work, we theoretically investigate how coherence can be extended by a sequence of quantum measurements on a diluted spin bath. In this regime, quantum measurement back-action plays a significant role: after each measurement, the quantum state of the environment is projected, and the distribution of possible values for the magnetic field created by the spin bath is narrowed, consequently extending the $T^*_2$ coherence time \cite{Cappellaro2012, Bonato2015, Bonato2017}. In other words, by learning the combined hyperfine coupling between the electron spin and the surrounding nuclei, one can infer the probability of the nuclear spin bath to be in a particular joint state. Thus, with every measurement performed on the central spin, additional information about its environment is gained. The learning process narrows down this probability distribution, thus reducing the fluctuations on the electron spin, leading to a longer coherence time. We will show that real-time adaptation of the measurement parameters, based on previous measurement outcomes, allows us to deterministically reduce the magnetic distribution to a narrow uni-modal distribution. In contrast to dynamical decoupling sequences, the improved coherence is maintained until the intrinsic quantum evolution due to inter-bath coupling broadens the set of possible coupling strengths. 

Our analysis starts with a discussion of the model employed to numerically simulate the central spin and the nuclear environment in Sec.~\ref{sec:model} and an introduction to the standard (non-adaptive) Ramsey measurement sequence in Sec.~\ref{sec:nonadapt}. We then introduce an adaptive learning protocol based on Bayesian estimation in Sec.~\ref{sec:adapt}, as well as assessing its ability to narrow spin bath distributions. Finally, in Sec~\ref{sec:focus}, we study the protocol's repeatability by simulating multiple intermittent narrowing sequences interleaved with free precession periods.
 
\section{Model}
\label{sec:model}
While the techniques studied here can be applied for more general systems with dilute environments, we focus on the electronic spin of a negatively charged nitrogen-vacancy (NV$^-$) centre in diamond that is coupled to a bath of $^{13}$C nuclear spins (Fig.~\ref{adaptive_schematic}).
NV$^-$ defects in diamond host an $S=1$ electronic spin that can be optically initialized and read-out. For simplicity, we assume the experiments are carried out at cryogenic temperatures, where fast initialization and single-shot read-out of the electronic spin can be performed with high fidelity \cite{Robledo2011}. We also assume that the concentration of electronic impurities (substitutional nitrogen) is sufficiently low that the main cause of decoherence is due to the hyperfine interaction with the surrounding $^{13}$C nuclear spins. The natural abundance of this isotope in diamond  is $\sim 1.1\%$ (but can be as low as $\sim 0.01\%$ for isotropically modified diamond samples). 

Following Refs.~\citen{Maze2008, Maze2012}, the Hamiltonian of our system can be written as:
\begin{equation}
H = H_{cs} + H_b + H^{int}_{cs-b}~,
\end{equation}
where $H_{cs}$, $H_{b}$ and $H^{int}_{cs-b}$ are the central spin (in our case the NV$^-$ electron spin) and spin bath Hamiltonians, and the interaction Hamiltonian of the defect and bath, respectively. In the presence of an external magnetic field $\mathbf{B} = (B_x,B_y,B_z)$, the individual Hamiltonian components can be written as
\begin{align}
\begin{split}
H_{cs} &= DS^2_z + \gamma_e \mathbf{B}\cdot\mathbf{S}~; \\
H_b &= \gamma_N \sum_n \mathbf{B}\cdot\mathbf{I}_n + \sum_{n<m} \mathbf{I}_n \cdot \mathbb{C}_{nm} \cdot \mathbf{I}_m~, \\
H^{int}_{cs-b} &= \sum_n \mathbf{S} \cdot \mathbb{A}_n \cdot \mathbf{I}_n~, 
\end{split}
\end{align}
where $D$ is the electron spin zero-field splitting, $\gamma_e$ and $\gamma_N$ are the electron and nuclear gyromagnetic ratios, respectively, $\mathbf{S} = (S_x, S_y, S_z)$ and $\mathbf{I}_n = (I_{n,x}, I_{n,y}, I_{n,z})$ are the spin operator vectors for the electron and $n^\mathrm{th}$ nuclear spin, respectively. Furthermore, $\mathbb{A}_n$ is the hyperfine tensor of the $n^\mathrm{th}$ spin, while $\mathbb{C}_{nm}$ is the coupling tensor between nuclei $n$ and $m$. Assuming the external field and NV$^-$ centre axis are aligned along the $z$-axis, we denote the Zeeman split states by $\{\Ket{\mu};~ \mu = -1,0,1 \}$. Moving to a rotating frame with respect to the electron $\mu = 0 \leftrightarrow 1$ transition, we obtain the Hamiltonian
\begin{align}
\begin{split}
H &= \sum^1_{\mu = 0} \Ket{\mu}\Bra{\mu} \otimes H_{\mu}~; \label{secular_ham}\\
H_{\mu} &= \sum_n \mathbf{\Omega}^{(\mu)}_n \cdot \mathbf{I}_n + \sum_{nm} \mathbf{I}_n \cdot \mathbb{C}^{(\mu)}_{nm} \cdot \mathbf{I}_m~,
\end{split}
\end{align}
where $\mathbf{\Omega}^{(\mu)}_n = \gamma_N \mathbf{B} + \mu \mathbf{A}_n$ [$\mathbf{A}_n = (A^{zx}_n,A^{zy}_n,A^{zz}_n)$ now denoting the hyperfine vector] is the effective Larmor vector for the $n^\mathrm{th}$ nuclear spin, and the form of $\mathbb{C}^{(\mu)}_{nm}$ can be found in \ref{app:bayesian}. We have also made the secular approximation, allowing the discarding of the zero-field splitting term. Given that in this work we consider a magnetic field along the $z$-axis, we measure the observable $A_z = \sum^N_{i=1} A^{zz}_i$, that is, the $z$ component of the hyperfine field of the nuclear ensemble. 

Full simulation of an ensemble of $N$ spins requires keeping track of $2^N$ complex elements of the density matrix, leading to an exponential scaling in computation time with Hilbert space dimension. For the case of a dilute bath, `clustering' approaches considerably reduce the computational time and make the problem tractable \cite{Maze2008, Witzel2012, Zhao2012b,Maze2012}. This stems from the idea that, due to the short-range nature of magnetic interaction, spin-spin coupling is important only within `clusters' of spins which happen to be very close to each other, while interaction between the clusters can be safely ignored. In our case, however, electron-nuclear interactions create correlations between the central electron spin and the nuclear spins. When the electron spin is measured, correlations are established between the clusters and their evolution cannot be considered as independent. As described in more detail in \ref{app:cluster}, this makes the clustering approach not applicable to our problem, and a full simulation of the density matrix evolution, for a limited number of nuclear spins, is required. In the following, we restrict our analysis to a full quantum simulation of a bath consisting of up to $10$ $^{13}$C nuclear spins.

Given a bath of $N$ nuclear spins around the central electron spin, the probability distribution $P(A_z)$ for the hyperfine interaction between the central spin and the bath can be computed as $P(A_z) = \{ \mathrm{Tr}(\Ket{A_i^z}\Bra{A_i^z} \rho_0):~ 1\leq i \leq 2^N \}$. Here $\rho_0$ is the initial spin bath density matrix, $\Ket{A_j^z}$ is the $j^\mathrm{th}$ joint spin bath eigenstate, with eigenvalue $A_j^z$ and $N$ is the number of nuclear spins in the environment. We assume the nuclear spin bath to be initially in the thermal state $\rho_0 = 2^{-N} \mathbb{I}_N$, which leads to a uniform $P(A_z)$. 
In this work, we take the ratio of the inverse of the standard deviation of the $A_z$ ($\sigma_z$) to it's original value ($\sigma_{z,0}$), which we shall refer to as the narrowing factor (N.F.) as a measure of how $T_2^*$ changes at each sequence, that is

\begin{equation} \label{T2est}
N.F. = \left(\frac{\sigma_z}{\sigma_{z,0}}\right)^{-1} = \frac{\left(\Braket{A^2_z}-\Braket{A_z}^2 \right)^{-1/2}}{\left(\Braket{A^2_z}_0-\Braket{A_z}_0^2 \right)^{-1/2}}~.
\end{equation}
where the zero subscript denotes a quantity taken with respect to the initial bath density matrix (that is, the completely mixed state), and $\Braket{A_z}$ and $\Braket{A^2_z}$ are the first and second moments of the distribution $P(A_z)$.

For simplicity, throughout this manuscript, we consider the case of perfect initialisation and (single-shot) readout fidelity for the electronic spin. State-of-the-art experiments on the NV centre in diamond at cryogenic temperature have reached initialization fidelities $>99\%$ and readout fidelities $>96\%$ \cite{Hensen2015}. A discussion of how adaptive measurements can be achieved in the absence of single-shot qubit readout is presented elsewhere \cite{Bonato2018}. As shown by previous experimental work \cite{Bonato2015}, small imperfections in qubit readout do not degrade the performance of adaptive protocols, as long as the number of repetitions is increased and the imperfections are correctly included when updating the knowledge about the system through Bayes rule. Furthermore, for reasons we explain in the following, we ignore any effects related to NV ionization and electron spin flips during readout (and their effect on the nuclear spin bath), as well as any changes in the hyperfine interaction when optically exciting the NV electronic spin. Single-shot electron spin readout is achieved by optical excitation of spin-selective transitions at cryogenic temperature \cite{Robledo2011b}. Due to excited-state mixing, this can result in electron spin flips that induce decoherence of the surrounding nuclear spins. This has been investigated in details by Reiserer \emph{et al} \cite{Reiserer2016}, who found that, while nuclear spin dephasing occurs after some hundred electron spin readouts, nuclear spin polarization is stable for thousands of readouts. In this work, we are only concerned about the polarization of the nuclear spins and not their phases, so we can safely neglect the effects of readout-induced electron spin flips on the nuclear spins.

\section{Non-adaptive Ramsey measurements}
\label{sec:nonadapt}
%

We consider a sequence of Ramsey measurements, i.e. interference experiments where the phase acquired by a spin under a magnetic field for a time $\tau$ is detected as a population difference between the spin eigenstates. The control parameters in a Ramsey experiment are the sensing time $\tau$ and the rotation angle $\phi$ of the detection basis. 

As discussed in Sec {\ref{sec:model}}, the nuclear spin bath initially induces a broad magnetic field distribution, corresponding to all possible configurations of an ensemble of $N$ nuclear spins ($I=1/2$, in our case).  Due to the interaction between the central spin and surrounding nuclear environment, the outcome from each Ramsey measurement provides partial information on the projection of the hyperfine field along the magnetic field axis (in our case, the $z$-axis). Through measurement back-action, the uncertainty in the bath state is changed with each measurement, depending on the measurement outcome. When the distribution is narrowed, as knowledge of the spin bath state increases, the uncertainty in the magnetic field felt by the central spin diminishes, thus increasing the $T^*_2$ coherence time. An example of this is shown in Fig.~\ref{bimodal}, where we plot the hyperfine distribution acting on the central electron spin after each  Ramsey experiment with $\tau_0 = 1 \upmu \mathrm{s}$ and $\phi = 0$. In general, a sequence of Ramsey experiment with fixed $\tau=\tau_0$ is not the optimal way to narrow the distribution. First, since measurement outcomes are random, one has no control over the final hyperfine distribution, and thus, the final state of the spin bath. This is illustrated in the example of Fig.~\ref{bimodal}: the final outcome of the process is a multi-modal four-peaked distribution (Fig.~\ref{bimodal}b) which corresponds to a $T^*_2$ of few $\upmu \mathrm{s}$ (Fig.~\ref{bimodal}c). Second, the sensitivity of Ramsey experiments is maximal when $\tau \sim T_2^*$. Given that $T_2^*$ is changed after each Ramsey experiment through measurement back-action, this suggests that one should adapt $\tau$ for each Ramsey measurement based on the current estimate of $T_2^*$. 

\section{Adaptive Bayesian protocol}
\label{sec:adapt}

In this section, we present an adaptive protocol that addresses the points evidenced above. Our protocol is based on Bayesian estimation, i.e. with each measurement, we update the classical probability distribution that represents our knowledge about the system and use it to extract the optimal parameters ($\tau$ and $\phi$) for the next sequence. This has two advantages: first, it maintains operation in the regime of maximum sensitivity for all steps in the protocol. Second, while measurement outcomes are random, online adaptation allows to select the best parameter settings to steer the distribution towards a narrow unimodal distribution . 

\begin{figure}[t!]
\centering
\includegraphics[width=\linewidth]{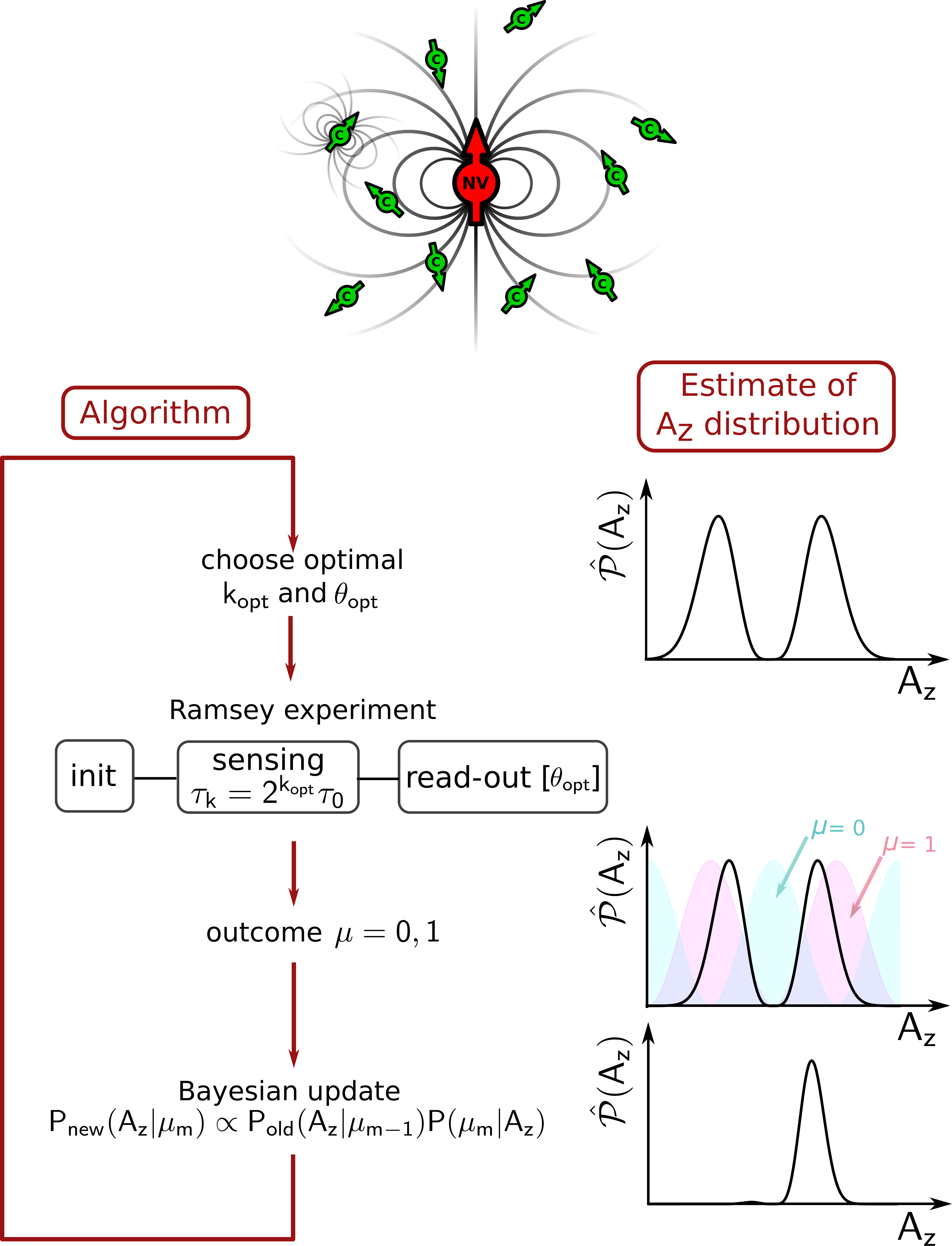}
\caption{
We consider a central electronic spin in a diluted nuclear spin bath (top). Our adaptive protocol chooses, in real-time, the optimal sensing time and phase for each Ramsey experiment with the goal to narrow the bath hyperfine interaction to a uni-modal distribution of minimal width. The algorithm keeps track of a classical probability distribution $P(A_z)$ to describe the bath, which is updated after each measurement by Bayes rule. $P(A_z)$ is used to estimate the optimal sensing time and phase.}
\label{adaptive_schematic}
\end{figure}

We now briefly describe the steps involved in the algorithm outlined in Fig.~\ref{adaptive_schematic}. Our knowledge about the bath is represented by a classical probability distribution $\hat{\mathcal{P}} (A_z)$, which approximates the true probability distribution $P(A_z)$. At the $m^ \mathrm{th}$ step,  $\hat{\mathcal{P}}(A_z)$ is updated using Bayes' theorem:
\begin{equation}\label{bayesianupdate}
\hat{\mathcal{P}}(A_z|\mu_1,...,\mu_{m}) = \hat{\mathcal{P}}(A_z|\mu_1,...,\mu_{m-1})\hat{\mathcal{P}}(\mu_{m}|A_z)~,
\end{equation}
\begin{figure*}[t!]
\centering
\includegraphics[width=.7\linewidth]{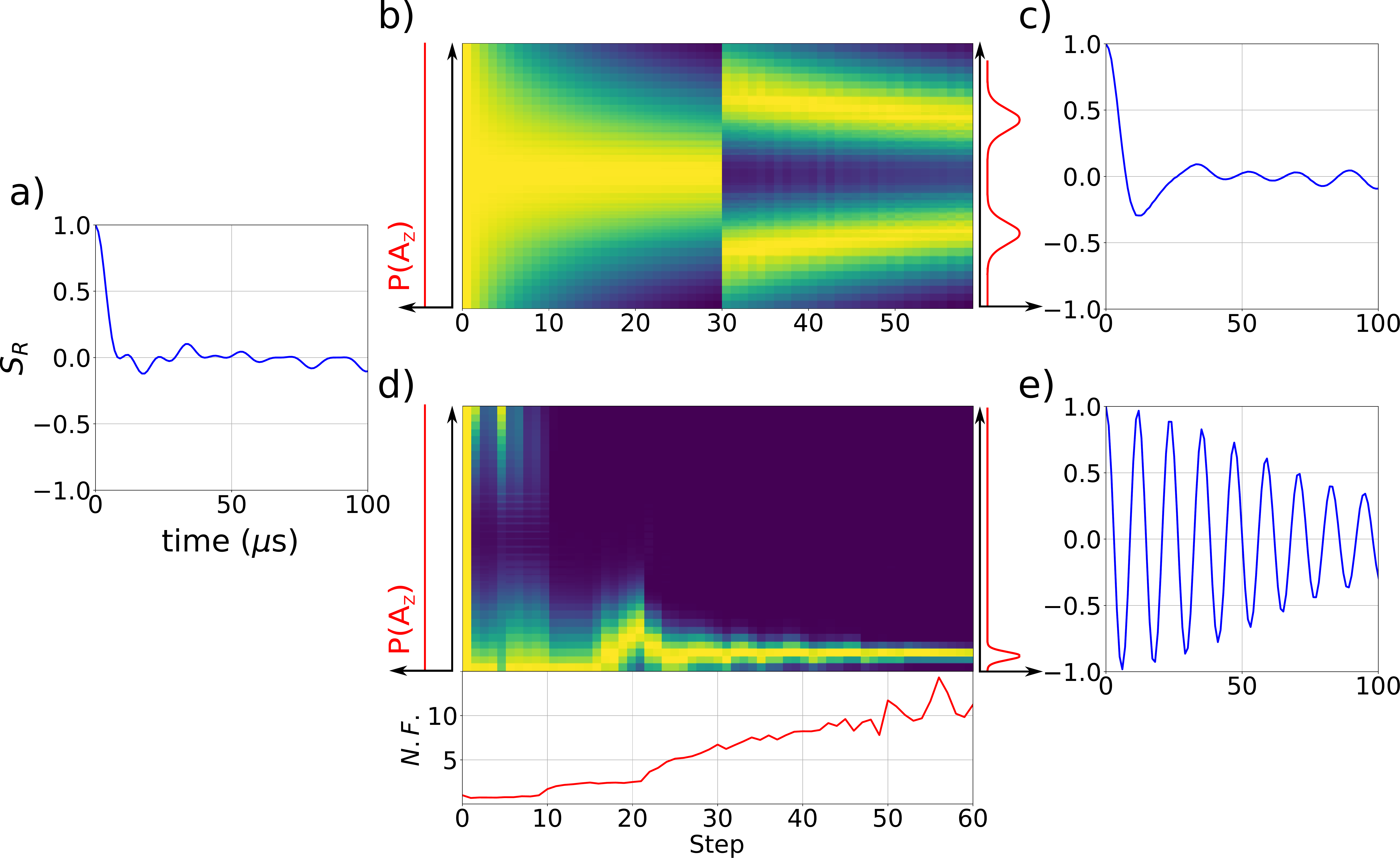}
\caption{
Evolution of the probability distribution $P(A_z)$ for the nuclear bath hyperfine interaction for a sequence of measurements on the electron spin, for the same initial bath distribution with a corresponding Ramsey signal (derived in \ref{app:ramsey}) shown in \textbf{a)}. \textbf{b)} For a sequence of non-adaptive Ramsey experiments, the resulting distribution is random, and can be multi-modal. \textbf{c)} The coherence time from the final Ramsey signal thus shows little improvement over the initial signal for the non-adaptive Ramsey scheme. \textbf{d)} Our adaptive protocol, on the other hand, selects the optimal parameters at each step, to deterministically narrow to a uni-modal distribution. \textbf{e)} Thus, the resulting Ramsey signal for the adaptive scheme shows a significant improvement in the coherence time. The red curves on the left and right of b) and c) are schematics of the initial and final probability distributions for the two Ramsey schemes, respectively. The red curve on the bottom shows the evolution of the narrowing factor for the adaptive scheme. Simulations were performed for a bath of 7 nuclear spins ($^{13}$C, $I=1/2$), under a magnetic field $B_z = 250$ mT.}
\label{bimodal}
\end{figure*}
where $\mu_l$ is the Ramsey result of the $l^\mathrm{th}$ estimation step (i.e. $\mu_l \in \{ 0, 1 \}$). The conditional probability $\hat{\mathcal{P}}(\mu_m|A_z)$ is given by \cite{Cappellaro2012, Bonato2017}
\begin{align}
\begin{split}
\hat{\mathcal{P}}(\mu_{m} = 0|A_z) &= \frac{1}{2} +\frac{1}{2} \mathrm{e}^{-(\tau / T^*_2)^2} \cos(2\pi A_z \tau + \phi) \nonumber\\
\hat{\mathcal{P}}(\mu_{m} = 1|A_z) &= 1 - \hat{\mathcal{P}}(\mu_{m} = 0|A_z)~,
\end{split}
\end{align}
where perfect readout fidelities for the two possible outcomes $\mu_{m} = 0$, and $\mu_{m} = 1$ are assumed.
\begin{figure}[t]
\begin{algorithm}[H]
\caption {Adaptive bath narrowing protocol}
\label{pseudo}
\begin{algorithmic}
\State $k = 0$ (\textit{sensing time index, $\tau_k = 2^k \tau_0$})
\\
\While{TRUE}
\\
\State calculate $k_{opt}$ - [Eq.~\eqref{k_opt}]
\\
\State calculate $\theta_{opt}$ - [Eq.~\eqref{phi_opt}]
\\
\State $\mu_k =$ Ramsey ($\tau = 2^k_{opt} \tau_0$, $\phi = \theta_{opt}$) 
\\
\If{$\mu_k \neq \mu_{k-1}$}
\State $\theta_{opt} \mathrel{+}= \frac{\pi}{2}$ 
\EndIf
\\
\State Bayesian update ($\mu$, $\tau = 2^k_{opt} \tau_0$, $\phi = \theta_{opt}$) -[Eq.~\eqref{bayesian_FT}]
[Ref.~\citen{Bonato2017}]
\\
\EndWhile
\end{algorithmic}
\end{algorithm}
\end{figure}
The probability distribution $\hat{\mathcal{P}}(A_z)$ can be expressed in Fourier space as \cite{Cappellaro2012, Bonato2015, Bonato2017}:
\begin{equation}
\hat{\mathcal{P}}(A_z) = \sum_j p_j \mathrm{e}^{i 2 \pi j A_z \tau_0}~.
\end{equation}
The $m^\mathrm{th}$ Bayesian update (Eq.~\eqref{bayesianupdate}) can be computed in terms of the Fourier coefficients $p_j$ as \cite{Cappellaro2012,Bonato2017}

\begin{align}
\begin{split}
&p^{(m)}_j = \frac{1}{2} p^{(m-1)}_j + \frac{1}{4} \mathrm{e}^{-(\tau_k / \hat{\mathcal{T}}^*_2)^2} \\
&\times\big[ \mathrm{e}^{i(\mu_m \pi + \phi^{(m)}_j)} p^{(m-1)}_{j-2^k} + \mathrm{e}^{-i(\mu_m \pi + \phi^{(m)}_k)} p^{(m-1)}_{j+2^k} \big]~, \label{bayesian_FT}
\end{split}
\end{align}
where $\phi^{(m)}_k$ is the angle of rotation of the detection basis. The current average hyperfine interaction $A^{avg}_z$ can be quantified in terms of a single Fourier coefficient, as
\begin{equation}
A^{avg}_z = \frac{1}{2 \pi \tau_0} \mathrm{arg}\Braket{\mathrm{e}^{i 2 \pi A_z \tau_0}} = \frac{1}{2 \pi \tau_0} \mathrm{arg}(p_{-1})~,
\end{equation}
where the last equality is a direct result of the Fourier representation of $\hat{\mathcal{P}}(A_z)$.

We choose the optimal sensing time based on the width of the current probability distribution $\hat{\mathcal{P}} (A_z)$ \cite{Bonato2015}, determined by its variance. For a well-defined, single mode distribution, the variance is equivalent to the \textit{Holevo variance} $V_H$, which can be described by a single $p_k$ coefficient \cite{Cappellaro2012}:

\begin{equation}\label{holevo}
V_H (\mathbf{A}_z) = \frac{1}{2} \left((2 \pi |p_1|)^{-2} - 1\right)~.
\end{equation}
This is very important for practical implementations since it allows choosing optimal sensing times with fast ($O(1)$ complexity) operations. Following previous work \cite{Cappellaro2012}, we restrict possible sensing times to the series $\tau_k = 2^k \tau_0$, where $\tau_0$ is the minimum sensing time. As described above, we heuristically choose the current sensing time to be as close as possible to the current estimated coherence time, which we denote by $\mathcal{T}_2^*$. 
Similar to to what we have done for the measured coherence time (see Eq.~\eqref{T2est}), we use the standard deviation of the distribution in order to monitor any changes to the estimated coherence time $\mathcal{T}_2^*$. Using the standard deviation associated with the Holevo variance $\sigma^H_z = V_H(A_z)^{1/2}$, we then choose a measurement time such that $\tau = 2^k \tau_0 \leq 1/ \sigma^{H}_z$, giving the current $k$ as: 

\begin{equation}\label{k_opt}
k_{opt} =  \left\lfloor\mathrm{log}_2 \left( \frac{\sigma^H_z}{2 \pi \tau_0} \right) + \mathrm{C} \right\rfloor~,
\end{equation}
where $\lfloor \cdot \rfloor$ denotes the closest smallest integer. The proportionality constant C here can be chosen to maximize the algorithm performance for the given specific experimental settings \cite{Bonato2017}.

\begin{figure}[t!]
\centering
\includegraphics[width=\linewidth]
{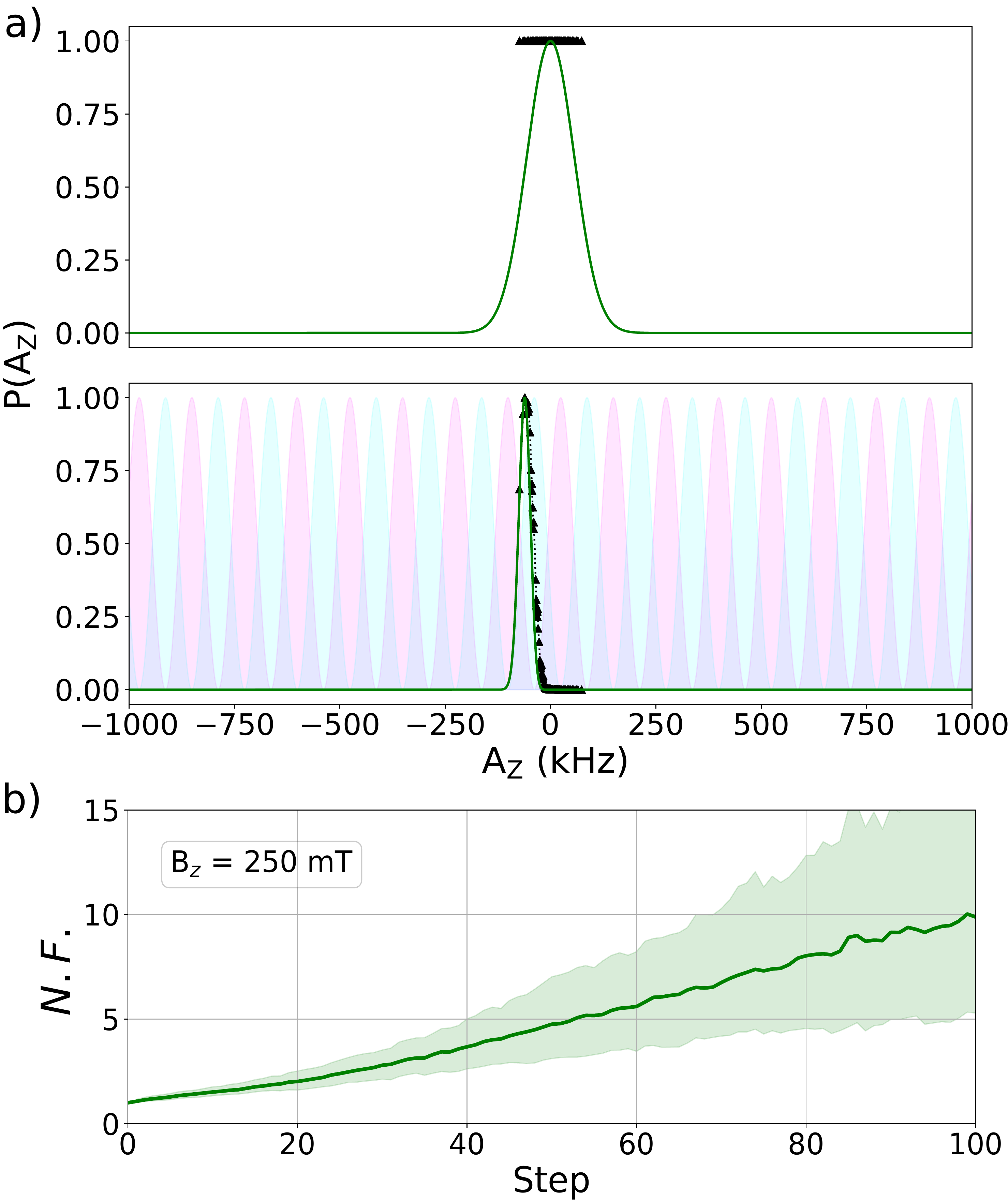}
\caption{\textbf{a)}Probability distribution $P(A_z)$ for a 7 nuclear spins with an applied magnetic field $B_z = 250 \mathrm{mT}$ before and after 20 Ramsey sequences (with $G=1$ and $F=0$). The green curve is the Bayesian distribution, whereas the discrete distribution in black is the distribution simulated directly from the Hamiltonian dynamics (showing the discretisation limited to $2^7=128$). The initial estimate is taken to be a normal distribution of width $\sim \tau^{-1}_0$. The conditional probabilities $P (\mu|A_z)$ for $\mu = 0$ and $\mu = 1$ are shown in magenta and cyan, respectively. \textbf{b)} The Bayesian narrowing scheme, showing the narrowing factor for averaged over 100 trials increasing with each additional step. In the above simulations, we chose $\tau_0 = 1 \upmu \mathrm{s}$, $G=3$ and $F=2$. The light shaded region is the standard deviation at each step.}
\label{narrow7spins}
\end{figure}


We implement $M_k = G + k F$ repetitions for the $k^\mathrm{th}$ Ramsey sequence ~\cite{Bonato2015, Higgins2009}, where $F$ and $G$ are integers, the latter being the number of repetitions of the shortest measurement sequence ($k=0$). For each Ramsey experiment, the rotation of the detection basis can locally optimized by finding the minimum of the Holevo variance in Fourier space \cite{Cappellaro2012,Bonato2017}:

\begin{equation}\label{phi_opt}
\phi^{(m)}_k = \theta_{opt} = \frac{1}{2} \mathrm{arg}(p_{-2^m})~.
\end{equation}
An outcome $\mu_i$  different than the previous outcome $\mu_{i-1}$ appears to often result in a bi-modal probability distribution $\hat{\mathcal{P}} (A_z)$. Through extensive simulations, we found that a unimodal distribution can be retrieved by including a conditional phase shift of $\pi/2$ in $\phi^{(m)}_k$, when the subsequent electron spin measurement result differs. The algorithm is described in details in Table ~\ref{pseudo}. The improvement over the non-adaptive algorithm is evident in Fig.~\ref{bimodal}, where we compare both algorithms on the same spin bath. While a sequence of non-adaptive Ramsey measurements results in a multi-modal distribution with a $T^*_2$ of a few $\upmu \mathrm{s}$, our adaptive protocol deterministically converges to a uni-modal distribution with $T^*_2 \sim 100 \upmu \mathrm{s}$. We note that, while this algorithm is based on a series of exponential sensing times, there are other possible strategies, such as sequential Monte Carlo protocols recently introduced for quantum sensing \cite{Santagati2019}. By using re-sampling strategies, these protocols may minimise the number of coefficients required in the Bayesian update, resulting in a more resource-efficient performance. In Fig.~\ref{narrow7spins}a, we give an example of how our scheme narrows the state probability distribution of bath of 7 spins over 20 Ramsey sequences, with repetition parameters $G = 3$ and $F = 2$. The applied magnetic field was set to $250 \mathrm{mT}$, with the shortest measurement time $\tau_0 = 1 \upmu \mathrm{s}$. 

\begin{figure}[t!]
\centering
\includegraphics[width=.8\linewidth]{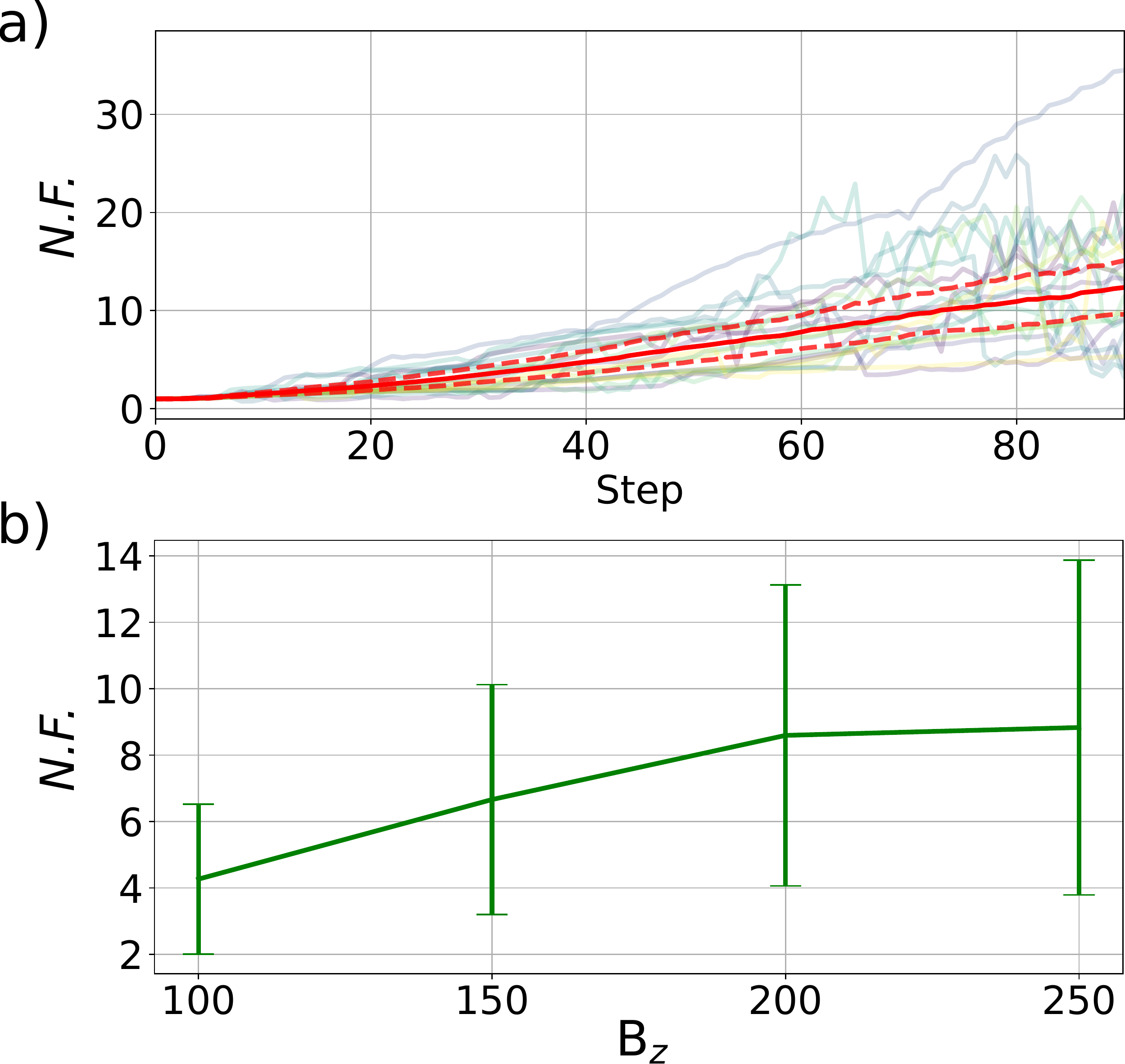}
\caption{\textbf{a)} Averaged narrowing factor (solid red) for an electron spin in a 7 spin nuclear environment with a magnetic of $B_z = 250 \mathrm{mT}$ and shortest measurement time $\tau_0 = 1 \upmu \mathrm{s}$, $G=3$ and $F=2$. 
In this case, the same spin environment was used for each realisation in order to show the different trajectories, determined by different random probabilistic measurement outcomes, are possible even when the central spin is surrounded by the same nuclear environment. 10 randomly selected trajectories (solid colours) from the 500 realisations are shown, as well as the standard deviation for each step of the narrowing algorithm (dashed red).
\textbf{b)} Magnetic field dependence of the final narrowing factor. For each magnetic field value, 100 random baths of 7 nuclear spins were generated (with $\tau_0 = 1 \upmu \mathrm{s}$, $G = 3$ and $F = 2$), averaging over the final narrowing factors. The saturation near $B_z = 250 \mathrm{mT}$ occurs primarily due to the finite discretisation of the simulated bath distribution, as discussed in the main text. Higher narrowing factors can be expected experimentally.}
\label{same_bath_Bdependence}
\end{figure}
%

In Fig.~\ref{narrow7spins}b, we plot the narrowing factor averaged over 100 different spin baths, simulated by randomly placing $^{13}$C spins in the diamond lattice. Our results show that, on average, the narrowing factor can be enhanced by at least a factor of 10.
In Fig~\ref{same_bath_Bdependence}a, we show how the narrowing algorithm increases the coherence time for a single spin bath for an applied magnetic field of $B_z = 250 \mathrm{mT}$ ($\tau_0 = 1 \upmu \mathrm{s}$, $G = 3$ and $F = 2$). Despite the fact that the same bath is used for each realisation of the algorithm, the probabilistic nature of the Ramsey experiment gives different optimal parameters (and thus, trajectories) for each realisation. In Fig.~\ref{same_bath_Bdependence}b, we show the dependence of the narrowing factor on the applied external magnetic field. As expected, as the magnetic field is increased, the spin bath fluctuations diminish as the hyperfine-field component parallel to the field (in this case, the $z$-component) is enhanced and dominates over the other components.

From Fig.~\ref{narrow7spins}b and Fig.~\ref{same_bath_Bdependence}b, we notice that the narrowing factor seems to saturate after a certain number of steps (Fig.~\ref{narrow7spins}b), or, alternatively as the applied magnetic field is increased (Fig.~\ref{same_bath_Bdependence}b). As we discuss presently, this saturation is merely a numerical artefact of the discrete nature of the simulated bath's probability distribution, from which we are extracting the narrowing factor. More specifically, for a bath comprising $N$ nuclear spins ($I=1/2$), the joint hyperfine eigenvalue $A_z$ can take $2^N$ values. Consequently, the  probability distribution describing the bath in the simulations is a discrete distribution consisting of $2^N$ points. This means that, for a small $N$, numerical issues may emerge when the distribution is narrowed significantly, up to the point where the distribution discretization is reached. It can therefore be expected that, when reaching the discretization limit, the standard deviation $\sigma_z$ would saturate. In other words, due to the finite size of the bath ($N=7$ in our simulations) we expect the narrowing to quickly reach a regime where only a few configurations of the bath are available,  thus limiting the validity of the simulated bath.
%
\begin{figure*}[t!]
\centering
\includegraphics[width=.6\linewidth]{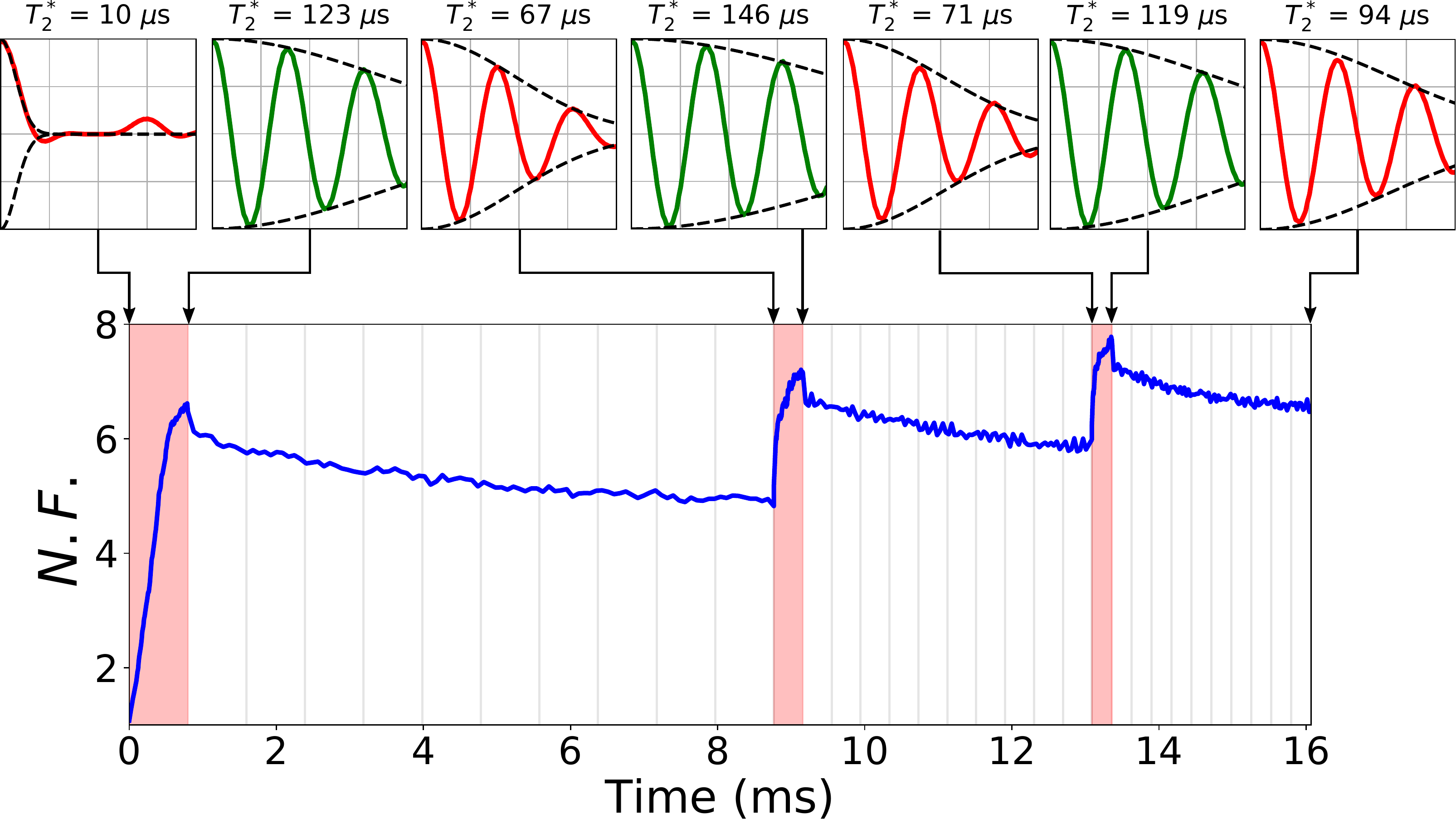}
\caption{After the bath distribution has been narrowed, it will broaden again over a timescale of the echo time $T_2$ due to inter-nuclear coupling. This effect can be mitigated by intermittent spin bath refocusing. In this example ($G=1$, $F=0$, $\tau_0 = 1 \upmu \mathrm{s}$) the bath is narrowed for about $1 \mathrm{ms}$, increasing $T^*_2$ from $10 \upmu \mathrm{s}$ to $123 \upmu \mathrm{s}$ (as shown by the Ramsey signals in the insets). The bath is then let freely evolve for 8ms, during which experiments can be performed on the electron spin exploiting the extended coherence. After $8 \mathrm{ms}$, $T^*_2$ is reduced to about $67 \upmu \mathrm{s}$. A second narrowing sequence (with the same adaptive algorithm) can be then performed, for a few hundred microseconds, which brings $T^*_2$ to $> 140 \upmu \mathrm{s}$. The whole sequence of narrowing steps and experiments can be repeated indefinitely, while maintaining long coherence. In the example, we cap the maximum narrowing factor to avoid reaching the discretisation of the probability distribution for a system of 7 nuclear spins.}
\label{refocus_4seqs}
\end{figure*}

\section{Spin bath refocussing}
\label{sec:focus}

Once the spin bath distribution has been narrowed, experiments can be performed with the extended coherence time $T_2^*$. This benefit would not, however, persist indefinitely since interactions between the nuclear spins induce diffusion and thus a broadening of $P(A_z)$ over a timescale of the echo time $T_2$.  
We consider the overall narrowing factor for multiple narrowing sequences back-to-back in order to asses the repeatability of the narrowing scheme. Remarkably, we find that it is indeed possible to refocus the spin bath even when each of the free precession periods last ten times as long as the preceding measurement sequence, as shown in Fig.~\ref{refocus_4seqs}, where we also show the Ramsey signal at several steps of the narrowing sequence for a single realisation, demonstrating the coherence time decay and subsequent revival to roughly 10 times its initial value. Due to to the exponential scaling of the Hilbert space, we restricted this simulation to a spin bath of 7 nuclear spins, although the same results can be obtained for larger environments, as we show in \ref{app:10spin} \footnote{In an experiment, this constraint would be lifted as we would only need to calculate the probability distribution based on experimental results, without the need simulate the computationally expensive spin environment.}.

\section{Conclusion}

In this paper, we have investigated a Bayesian adaptive approach to environment state learning. In recent work, such an approach has been shown to reach the quantum limit of parameter estimation \cite{Cappellaro2012}. We have applied a modified version of this algorithm to gain information about the dilute nuclear environment of a solid state nanostructure (an example of which would be an NV$^-$ centre), in the presence of a known, external magnetic field. We have shown that, in turn, this information reduces the magnetic field fluctuations felt by the central spin, as the state of the surrounding nuclear ensemble is partially projected and steered towards a random, but particular, state.
Whilst focussing on the specific environment of NV$^-$ centres, these results can be straightforwardly extended to other systems having sparse nuclear spin environments. 

Several interesting applications of this technique can be investigated in the near future. An exciting extension of the narrowing scheme would be to engineer exotic spin bath states. The full control over the measurement times and readout angle means that it is possible, in principle, to `guide' the spin bath into any state by choosing, instead of optimising, the measurement time and phase. Thus, by utilising this partial projection of the spin bath distribution, one can realise interesting many-body states which may show, for example, different entanglement properties and other interesting features.

\section{Acknowledgements}
We thank Jeronimo Maze, Hossein Dinani and Dominic Berry for helpful discussions. This work is supported by the Engineering and Physical Sciences Research Council (grants EP/P019803/1, EP/S000550/1), the European Commission (QuanTELCO, project ID:862721) and the Carnegie Trust for Scotland (Research Incentive Grant RIG007503). EMG acknowledges support from the Royal Society of Edinburgh and the Scottish Government.

\section*{Appendix}
\appendix
\section{Secular approximation correction}
\label{app:bayesian}

In Sec.~\ref{sec:model}, we performed the secular approximation for high fields, which allowed us to arrive at the simple form of the Hamiltonian, given by Eq.~\eqref{secular_ham}. In order to obtain Eq.~\eqref{secular_ham}, however, the components of the hyperfine fields from $^{13}$C orthogonal to the NV axis have to be accounted for perturbatively, as these terms may lead to quantitative changes in the dynamics of the spin bath evolution \cite{Childress2006, Maze2008}. Following Refs.~\citen{Childress2006, Maze2008}, with the magnetic field parallel to the $z$-axis, we get an effective nuclear-nuclear spin coupling for the $n^\mathrm{th}$ Ramsey experiment given by 

\begin{equation}
\mathbb{C}^{eff}_{ij} = \mathbb{C}_{ij} + \delta\mathbb{C}^{(\mu_n)}_{ij}~,
\end{equation}
where $\mathbb{C}^{(\mu_n)}_{ij}$ is the correction to the interaction term between the $i^\mathrm{th}$ and $j^\mathrm{th}$ nuclear spins, and is given by

\begin{equation}
\delta\mathbb{C}^{(\mu_n)}_{ij} = -\frac{D( \gamma_N / \gamma_e)^2}{2-3 \mu_n} \delta \mathbf{g}^T_i \cdot \delta \mathbf{g}_j~,
\end{equation}
where $\mathbf{g}_i$ is the correction term to the $i^\mathrm{th}$ spin g-tensor, which captures the effects of the additional hyperfine transverse terms:

\begin{equation}
\delta \mathbf{g}_i = \frac{(2-3 \mu_n) \gamma_e}{D \gamma_N}\left(\begin{array}{cccc} A^{xx}_n &  A^{xy}_n &  A^{xz}_n \\ 
A^{yx}_n &  A^{yy}_n &  A^{yz}_n \\ 
0 & 0 & 0 \end{array}\right)~.
\end{equation}

\section{Invalidity of cluster algorithm}
\label{app:cluster}

As discussed in Sec.~\ref{sec:model}, the cluster approximation for a diluted bath cannot be used in the case of correlated measurements. Indeed, consider the density matrix after the $n^\mathrm{th}$ Ramsey experiment, $\rho^{full}_n$, which is obtained by propagating the density matrix after the $(n-1)^\mathrm{th}$ step:

\begin{equation}\label{full}
\rho^{full}_n = \mathcal{U}_{\mu_n} \rho^{full}_{n-1} \mathcal{U}^\dagger_{\mu_n}~,
\end{equation}
where $\mathcal{U}_{\mu_n} = \bigotimes^{N}_{k=1} U^{(k)}_0 + (-1)^{\mu_n +1}\bigotimes^{N}_{k=1} U^{(k)}_1 $, and 

\begin{align}
\begin{split}
U^{(k)}_{\mu_n} &= \exp\left[-i \frac{t}{\hbar} H^{(k)}_{\mu_n} \right] \\
&= \exp\left[-i \frac{t}{\hbar}\left(\mathbf{\Omega}^{(\mu)}_k \cdot \mathbf{I}_k + \sum_m \mathbf{I}_k \cdot \mathbb{C}^{(\mu)}_{km} \cdot \mathbf{I}_m \right) \right]~,
\end{split}
\end{align}
is the time evolution operator of the spin bath given a $\mu_n = 0,1$ Ramsey experiment result. The cluster approximation assumes that the nuclear environment can be classified into smaller clusters of spins such that the density matrix $\rho^{full}_{n-1}$ can be broken down into a product of smaller density matrices $\bigotimes^{N_c}_{j=1} \rho^{(j)}_{n-1}$, where $N_c$ is the number of clusters and each $\rho^{(j)}_{n-1}$ is the density matrix for a cluster of spins. Furthermore, the time evolution operator can also be clustered as $\mathcal{U}_{\mu_n} = \bigotimes^{N_c}_{j=1} U^{(j)}_0 + (-1)^{\mu_n +1} \bigotimes^{N_c}_{j=1} U^{(j)}_1$, where $U^{(j)}_{\mu_n}$ now describes the time evolution of the $j^\mathrm{th}$ cluster. Then, the full density matrix in Eq.~\eqref{full} can be expanded as

\begin{align}
\begin{split}
\rho^{full}_n = &\bigotimes^{N_c}_{j=1} U^{(j)}_0 \rho^{(j)}_{n-1} U^{(j)\dagger}_0 + \bigotimes^{N_c}_{j=1} U^{(j)}_1 \rho^{(j)}_{n-1} U^{(j)\dagger}_1 \\
+ &\bigotimes^{N_c}_{j=1} U^{(j)}_1 \rho^{(j)}_{n-1} U^{(j)\dagger}_0 + \bigotimes^{N_c}_{j=1} U^{(j)}_0 \rho^{(j)}_{n-1} U^{(j)\dagger}_1~, \label{good}
\end{split}
\end{align}
since intercluster nuclear-nuclear interactions are negligible and can thus be ignored as long as a reasonable number of clusters $N_c$ is taken \cite{Maze2008}. We have taken the measurement outcome to be $\mu_n = 1$, without loss of generality. From the above, we cannot obtain each individual cluster density matrix since we cannot factor the above as a single Kronecker product over clusters. If the cluster approximation holds, then we should be able to obtain the same full density matrix by propagating each individual cluster density matrix separately, and then taking the Kronecker product, that is

\begin{align}
\begin{split}
&\rho^{clus}_n = \bigotimes^{N_c}_{j=1} \rho^{(j)}_{n-1}~; \\
&\rho^{(j)}_{n-1} = \mathcal{U}^{(j)}_{\mu_n} \rho^{(j)}_{n-1}  \mathcal{U}^{(j)\dagger}_{\mu_n}~,
\end{split}
\end{align}
where $\mathcal{U}^{(j)}_{\mu_n} = \bigotimes^{N_j}_{k=1} U^{(k)}_0 + (-1)^{\mu_n +1}\bigotimes^{N_j}_{k=1} U^{(k)}_1 $, with $N_j$ being the number of spins in the $j^\mathrm{th}$ cluster. $\rho^{clus}_n$ can then be written down as

\begin{align}
\begin{split}
\rho^{clus}_n = \bigotimes_k &\left( U^{(k)}_0  \rho^{(k)}_{n-1} U^{(k)\dagger}_0 + U^{(k)}_1 \rho^{(k)}_{n-1} U^{(k)\dagger}_1 \right.\\
&\left.+ U^{(k)}_1 \rho^{(k)}_{n-1} U^{(k)\dagger}_0 + U^{(k)}_0 \rho^{(k)}_{n-1} U^{(k)\dagger}_1 \right)~, \label{bad}
\end{split}
\end{align}
thus giving a different (and incorrect) expression for the density matrix of the spin bath at step $n$ since $\rho^{clus} \neq \rho^{full}$.

\section{Ramsey signal}
\label{app:ramsey}

Starting with the standard Ramsey sequence $R_x \left( \frac{\pi}{2} \right) - U(\tau) - R_x \left( \frac{\pi}{2} \right)$, with $R_x(\theta)$ and $U(\tau)$ denoting a $\theta$-rotation about the $x$ axis and a free precession for a time $\tau$ respectively, the Ramsey signal can be calculated as

\begin{align}\label{ramsey_sig}
\begin{split}
s(\tau) &=\mathrm{Tr}_\mathrm{bath}\{\langle 0|\rho(\tau)| 0\rangle \}~, \\ 
\rho(\tau) &=U_{\mathrm{Ram}}(\tau) \left[\rho_{\mathrm{e}} \otimes \rho_\mathrm{bath} \right] U_{\mathrm{Ram}}(\tau)^{\dagger} ~,
\end{split}
\end{align}
where $U_{\mathrm{Ram}}(\tau) = R_x \left( \frac{\pi}{2} \right) U(\tau) R_x \left( \frac{\pi}{2} \right)$, and $\rho_{\mathrm{e}}$ and $\rho_\mathrm{bath}$ are the electron and bath density matrices, respectively. By simplifying Eq.~\eqref{ramsey_sig}, we finally obtain

\begin{equation}
s(\tau) = \frac{1}{2}\Big[1- \mathrm{Tr} \{ U_0 (\tau) \rho_\mathrm{bath} U^\dagger_1 (\tau) \} \Big] = \frac{1}{2}\Big[1-S_R(\tau) \Big]~,
\end{equation}
where $U_{\mu \in \{0, 1\}} (\tau)  = \bigotimes^N_{k=1} U^{(k)}_\mu (\tau)$ defined in \ref{app:cluster}, and we refer to the time-varying component $S_R(\tau) = \mathrm{Tr} \{ U_0 (\tau) \rho_\mathrm{bath} U^\dagger_1 (\tau) \}$ as the Ramsey signal shown in Fig.~\ref{bimodal} and Fig.~\ref{refocus_4seqs} in the main text. For the latter, the coherence time $T^*_2$ is extracted by fitting the envelope of the signal with $\exp\left[ -(\tau / T^*_2)^2 \right]$, with any small deviations from the signal stemming from the subsequent revivals of the signal.

\section{10 spin bath results}
\label{app:10spin}

\begin{figure}[t!]
\centering
\includegraphics[width=\linewidth]{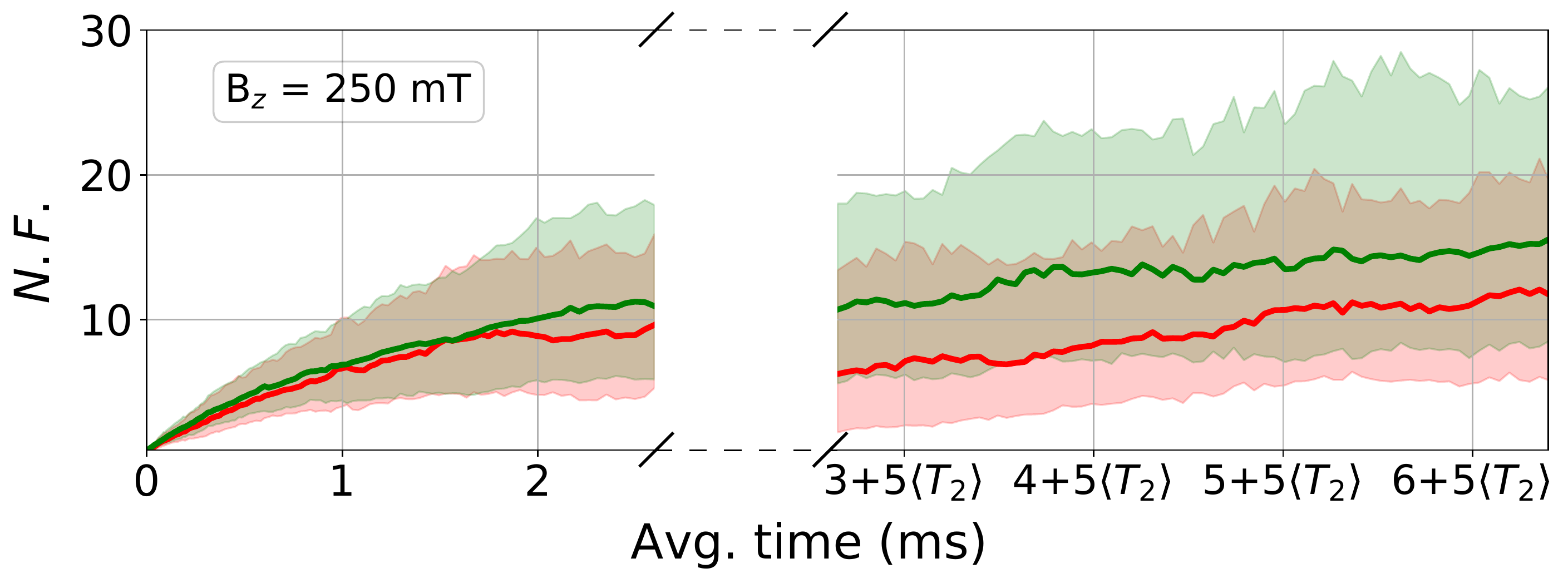}
\caption{Comparison of average narrowing sequence applied to a  10 spin nuclear environment with continuous narrowing (green), and with a free precession period of $\sim 5 \langle T_2 \rangle$ (red) represented by the blue shaded region, where $\langle T_2 \rangle \sim 100 \mathrm{ms}$ is the $T_2$ time averaged over all the runs. Whilst, as expected, the spin bath diffusion during between narrowing sequences reduces the narrowing fraction considerably, the `intermittent' case slowly approaches the continuous narrowing curve as we approach the saturation $T^*_2$ (limited by the $T_2$ time). The end of the first narrowing period for each experiment realisation was aligned for better graphical representation of the narrowing factor decrement.}
\label{app:refocus}
\end{figure}

Due to the exponential scaling of the spin bath density matrix, we have restricted ourselves for baths of 7 spins in the main text. However, we are also able to obtain similar results for larger numbers of spins, as we show in Fig.~\ref{app:refocus}, where we compare the narrowing factor for the `interrupted' sequence with the factor obtained by continuously applying the narrowing sequence, with the gap denoting the period during which the experimenter may make use of the electron spin. 

\clearpage
\bibliography{Bayesian}
\bibliographystyle{unsrt}

\end{document}